\documentclass[conference]{IEEEtran}
\IEEEoverridecommandlockouts
\usepackage{cite}
\usepackage{amsmath,amssymb,amsfonts}
\usepackage{algorithmic}
\usepackage{graphicx}
\usepackage{textcomp}
\usepackage{xcolor}
\usepackage{subfigure}
\usepackage{multirow}
\def\BibTeX{{\rm B\kern-.05em{\sc i\kern-.025em b}\kern-.08em
    T\kern-.1667em\lower.7ex\hbox{E}\kern-.125emX}}
\begin{document}

\title{A Discourse-level Multi-scale Prosodic Model for Fine-grained Emotion Analysis\\
% {\footnotesize \textsuperscript{*}Note: Sub-titles are not captured in Xplore and
% should not be used}
% \thanks{Identify applicable funding agency here. If none, delete this.}
}

\author{\IEEEauthorblockN{Xianhao Wei}
\IEEEauthorblockA{\textit{Department of Computer Science and Technology} \\
\textit{Tsinghua University}\\
Beijing, China \\
weixh20@mails.tsinghua.edu.cn}
\and
\IEEEauthorblockN{Jia Jia$^{*}$\thanks{* Corresponding author.}}
\IEEEauthorblockA{\textit{Department of Computer Science and Technology} \\
\textit{Tsinghua University}\\
Beijing, China \\
jjia@tsinghua.edu.cn}
\and
\IEEEauthorblockN{Xiang Li}
\IEEEauthorblockA{\textit{Shenzhen International Graduate School} \\
\textit{Tsinghua University}\\
Shenzhen, China \\
xiang-li20@mails.tsinghua.edu.cn}
\and
\IEEEauthorblockN{Zhiyong Wu}
\IEEEauthorblockA{\textit{Shenzhen International Graduate School} \\
\textit{Tsinghua University}\\
Shenzhen, China \\
zywu@sz.tsinghua.edu.cn}
\and
\IEEEauthorblockN{Ziyi Wang}
\IEEEauthorblockA{\textit{Academy of Arts and Design} \\
\textit{Tsinghua University}\\
Beijing, China \\
wzyzoeywang@gmail.com}
% \and
% \IEEEauthorblockN{6\textsuperscript{th} Given Name Surname}
% \IEEEauthorblockA{\textit{dept. name of organization (of Aff.)} \\
% \textit{name of organization (of Aff.)}\\
% City, Country \\
% email address or ORCID}
% }
}

\maketitle

\begin{abstract}
This paper explores predicting suitable prosodic features for fine-grained emotion analysis from the discourse-level text. To obtain fine-grained emotional prosodic features as predictive values for our model, we extract a phoneme-level Local Prosody Embedding sequence (LPEs) and a Global Style Embedding as prosodic speech features from the speech with the help of a style transfer model. We propose a Discourse-level Multi-scale text Prosodic Model (D-MPM) that exploits multi-scale text to predict these two prosodic features. The proposed model can be used to analyze better emotional prosodic features and thus guide the speech synthesis model to synthesize more expressive speech. To quantitatively evaluate the proposed model, we contribute a new and large-scale Discourse-level Chinese Audiobook (DCA) dataset with more than 13,000 utterances annotated sequences to evaluate the proposed model. Experimental results on the DCA dataset show that the multi-scale text information effectively helps to predict prosodic features, and the discourse-level text improves both the overall coherence and the user experience. 
More interestingly, although we aim at the synthesis effect of the style transfer model, the synthesized speech by the proposed text prosodic analysis model is even better than the style transfer from the original speech in some user evaluation indicators.

\end{abstract}

\begin{IEEEkeywords}
Local Prosody Embedding sequence, discourse analysis, prosody, multi-scale, fine-grained
\end{IEEEkeywords}

\section{Introduction}
In the busy modern life, people are under more and more pressure, and many people relax by reading. However, as everyday life squeezes people's limited time, the gap between the need to read and people's free time is widening. In this context, people are more and more inclined to choose AudioBooks. High-quality AudioBooks often require tremendous cost support, which is time-consuming and labor-intensive. The development and maturity of speech synthesis technology bring new opportunities and challenges to the application of AudioBooks.

In recent years, end-to-end text-to-speech (TTS) models such as Tacotron2~\cite{JonathanShen2018NaturalTS} have been able to synthesize speech sound that closely resembles a human voice. However, correctly pronouncing every word is not enough to tell a story vividly. It often requires synthesizing more expressive speech, rather than mechanically pronouncing every word, to keep the listeners interested. The traditional approach to control the style of synthesized speech is usually to extract a style embedding from reference speech, which is then used to synthesize target speech of the same style \cite{RJSkerryRyan2018TowardsEP,YuxuanWang2018StyleTU,DaisyStanton2018PredictingES,RafaelValle2020MellotronME,MattWhitehill2020MultiReferenceNT}. Synthesizing a discourse-level story presents some new challenges. On the one hand, today's speech synthesis systems usually only consider one single utterance, and there is no connection between utterances. However, a story contains more than one utterance, and the context may influence each utterance's most appropriate prosody style. On the other hand, for a long time, almost all researches focused on how to transfer style from reference speech better \cite{RJSkerryRyan2018TowardsEP,SiddharthGururani2019ProsodyTI}, without paying attention to whether the style is suitable for the text. Even if in some research \cite{DaisyStanton2018PredictingES,YiLei2021FineGrainedES} style embedding can be predicted from the text during inference, this design only regards the text analysis module as an accessory rather than a carefully designed one. They all ignore the high correlation between the textual and acoustic modalities. This fact is more crucial in long speech synthesis tasks such as AudioBooks. If the prosody of the synthesized speech does not match the content of the text, it significantly detracts from the listeners' experience. For AudioBooks, the key is on how to automatically analyze the most appropriate prosody features from the text for the speech, rather than transfer style from reference speech.

To address the aforementioned problems, we have constructed a Discourse-level Chinese AudioBooks (DCA) dataset. It is worth noting that the utterances in our dataset are semantically contiguous. We construct a Discourse-level Multi-scale Prosodic analysis Model (D-MPM) based on this dataset to support stylized speech synthesis. The model uses the text information of multiple scales, controls the overall style of synthesized speech by predicting a global style embedding (GSE), and controls local prosody by predicting local prosody embedding (LPE) on the fine-grained phoneme level.

\begin{figure*}
    \centering
    \includegraphics[width=\textwidth]{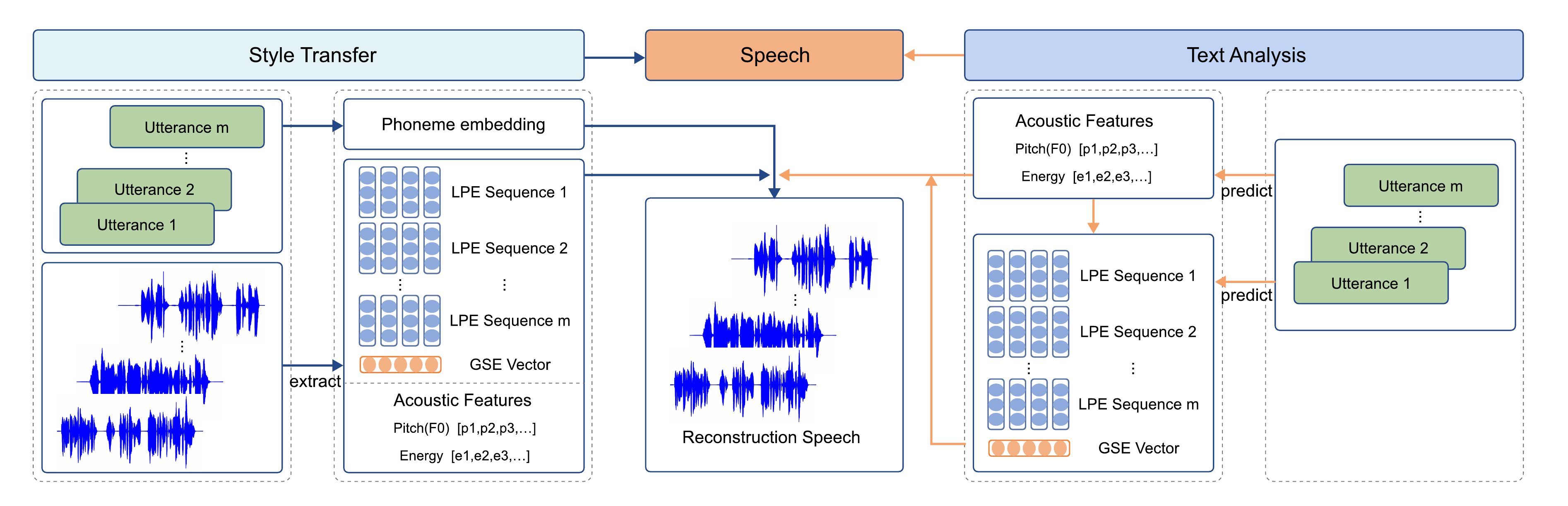}
    \caption{The overall system framework.}
    \label{fig:pipeline}
\end{figure*}

We first construct a discourse-level style transfer system similar to~\cite{XiangLi2021TowardsMS}. The system extracts phone-level fine-grained LPE sequences independent of the specific pronunciation from the reference speech to control the prosody of the synthesized speech. In addition, the model also extracts a GSE for each discourse to control the global style. Finally, both LPE and GSE are used to reconstruct speech based on Tacotron2~\cite{JonathanShen2018NaturalTS}. However, wo do not focus on the style transfer system but a well-designed discourse-level text prosody analysis model. As mentioned above, when synthesizing speech, using reference speech will lead to a fragmentation of text content and style, and in many cases, there is no reference speech at all. The proposed D-MPM can effectively bridge this shortcoming and predict both LPE and GSE from the text, making the predicted features more consistent with the text content.

To summarize, this work makes the following three main contributions:
\begin{itemize}
    \item[$\bullet$] We have built a benchmark dataset of high-quality Discourse-level Chinese AudioBook (DCA) dataset, including multiple speakers and styles, and annotated with pinyin (Chinese pronunciation notation) and tone information of corresponding texts. Subsequent additional utterance-level emotional information is still in the labeling process. We will make all these data and annotations public in the future.
    \item[$\bullet$] We construct a Discourse-level Multi-scale Prosodic analysis Model (D-MPM) to support stylized speech synthesis. Unlike traditional single-utterance analysis and synthetic models, our model can acquire knowledge from a richer context to better use the hierarchical relationship of discourse-level data.
    \item[$\bullet$] We provide a framework for predicting fine-grained phoneme-level prosodic features from text. We fuse multi-scale text features together with some predicted acoustic features to achieve automatic fine-grained control of speech prosody.
\end{itemize}

\section{Related Work}
% \subsection{Expressive TTS}
Expressive TTS is designed to synthesize highly expressive speech. Early researches mainly control speech style through style labels \cite{HieuThiLuong2017AdaptingAC,NicolasObin2014SLAMAS,JaimeLorenzoTrueba2018InvestigatingDR}. This form is expensive for data labeling, and this supervised method also has higher quality requirements for data labeling. Skerry-Ryan et al.~\cite{RJSkerryRyan2018TowardsEP} proposed a reference encoder to extract style features from reference speech. This model illustrates that the style of speech can be modeled and controlled in ways such as style embedding. However, this method requires reference speech. The speech synthesized by this method can only be consistent with the style of the reference speech but cannot guarantee that its style matches the text. Earlier studies only used a style embedding like global style token (GST)~\cite{YuxuanWang2018StyleTU} to control the synthesized style, which is insufficient for fine-grained control. Some researchers \cite{YounggunLee2019RobustAF,GuangzhiSun2020FullyHierarchicalFP,YiLei2021FineGrainedES} have shown that using more fine-grained features to control the style can make the prosody within the utterance more vivid and thus have higher expressiveness. However, fine-grained features are highly correlated with speech length, leading to transfer learning difficulties.

Many approaches not only support transfer learning from reference speech but also use text analysis modules to predict style information during inference \cite{DaisyStanton2018PredictingES,MaxMorrison2020ControllableNP,QicongXie2021MultispeakerMT}. This strategy solves the problem of no reference speech during inference and also makes the synthesized speech relevant to the textual content to a certain extent. However, the text analysis module in almost all models is just a simple adjunct to the overall speech synthesis system. A common practice is to use the output of the text encoder of the speech synthesis module to predict style embedding. Phonemes are usually used as text input because the correspondence between phonemes and sounds is closer, making it easier for the model to learn pronunciation information. However, predicting style features of corresponding speech from text is inherently a natural language understanding (NLU) task. This kind of problem usually uses the word as input. Fine-grained textual information like phonemes can make it more difficult for the model to understand the content of the text, especially in a language like Chinese. This contradiction makes the previous practice of using the output of the Text Encoder of the synthesis module to predict the style control information to a large extent limiting the ability to understand the text content. \cite{YukiyaHono2020HierarchicalMG} use a hierarchical model and multi-scale (Word, Phrase, and Utterance) VAE to control the style, exploiting the relationship between multiple scale of text. \cite{SercanOArik2017DeepVR,AdrianLancucki2021FastpitchPT,YiRen2021FastSpeech2F} introduced the prediction of acoustic features in the process of text-to-speech task, which made the prosody of the synthesized speech more abundant.

% \subsection{Pretrained model for NLU tasks}
In the NLU tasks, pre-trained models achieve state-of-the-art performance across many subtasks \cite{radford2018improving,JacobDevlin2018BERTPO,MatthewEPeters2018DeepCW,cui-etal-2020-revisiting}. These models are trained unsupervised on a large amount of data in advance and can represent the text feature well. Compared with traditional models such as CNN and RNN, pre-trained models have more prior knowledge. This makes fine-tuning based on pre-trained models often result in better performance. Discourse-level data can effectively improve the performance of reading comprehension~\cite{TodorMihaylov2019DiscourseAwareSS}. Step-by-step fusion of low-level features into high-level features using a hierarchical model is a more general approach to dealing with discourse-level input \cite{DuyuTang2015DocumentMW,StefanosAngelidis2018MultipleIL}. In this work, we propose a well-designed discourse-level text analysis model. It takes the respective advantages of different scale of text to fully mine the mapping relationship between textual features and acoustic features. In addition to fusing low-level features into high-level ones, our model also adds a feedback mechanism to feedback high-level fused features to low-level features directly.

\section{Problem Formulation}
A speech synthesis system can be viewed as a mapping from text to acoustic modality. We focus on the prosody style features of the synthesized speech, hoping to synthesize a highly expressive speech. For one utterance, the style transfer approach is to learn the corresponding prosody features from the reference speech and use these features to synthesize the target speech. Text analysis aims to analyze the corresponding prosody directly from the text without any reference speech.

Our system, shown in Figure~\ref{fig:pipeline}, consists of two main components: (1) A speech style transfer part for modeling the hidden prosodic representation of speech; (2) A discourse-level multi-scale text prosodic analysis model, which targets the prosodic features extracted from the first component. It can be used to replace the prosodic features extracted from speech to achieve automatic expressive speech synthesis in line with text features.

To make the problem clearer, we define our problem as follows: given a piece of discourse-level text $D = (U_{1}, U_{2}, ... , U_{m})$ containing $m$ distinct consecutive utterances. For an utterance $U_{i}$ with $p_{i}$ phonemes, we need to predict the multi-scale acoustic features:
\begin{equation}
    f: (D : U_{1}, U_{2}, ... , U_{m}) \Rightarrow (y : y_{gse}, y_{lpe}, y_{pitch}, y_{energy})
\end{equation}

\begin{figure*}
    \centering
    \includegraphics[width=\textwidth]{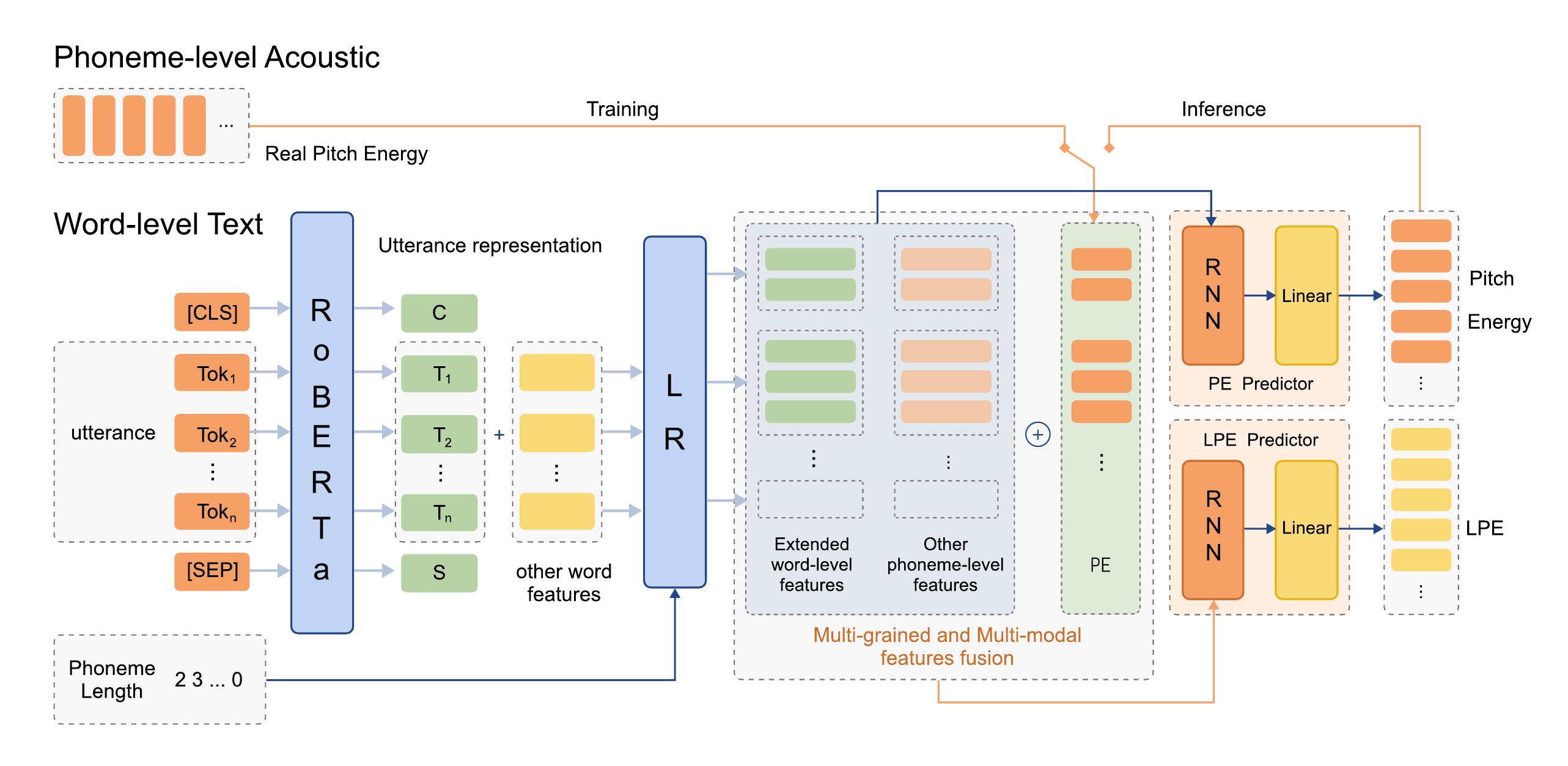}
    \caption{A flowchart of the processing of a single utterance by our proposed U-MPM. It is the first stage of our model. LR refers to the Length Regulator module, and it can extend the word-level embeddings to the phoneme level according to the phoneme length of each word. During training, the model predicts acoustic features (pitch and energy) from textual features and concatenates textual features and ground-truth acoustic features to predict local prosody embedding (LPE). During inference, the predicted acoustic features are used in place of the ground truth. $\oplus$ denotes the concatenation operator.}
    \label{fig:3M_sentence}
\end{figure*}

\section{Methodology}
We first perform feature extraction on the discourse-level speech data based on a trained speech style transfer system and use the extracted features to reconstruct the speech. Then these acoustic features are predicted using D-MPM for the automatic discourse-based synthesis of AudioBooks. This section is divided into three parts: 1) Briefly describe how we process data and extract features from speech; 2) Utilize our proposed Utterance-level Multi-scale Prosodic analysis Model (U-MPM) to predict local prosody embedding (LPE); 3) Based on the second part, we use the discourse-level data to construct our Discourse-level Multi-scale Prosodic analysis Model (D-MPM) to enrich the LPEs features and predict the overall global style embedding (GSE). Since there are only two styles in our DCA dataset, we predict style labels instead of directly predicting GSE, and the impact of GSE is relatively small. This paper mainly discusses the impact of LPEs and the GSE mentioned in the sequel all refer to style labels.

\subsection{Data Processing and Feature Extraction}
\begin{figure}
    \centering
    \includegraphics[width=\columnwidth]{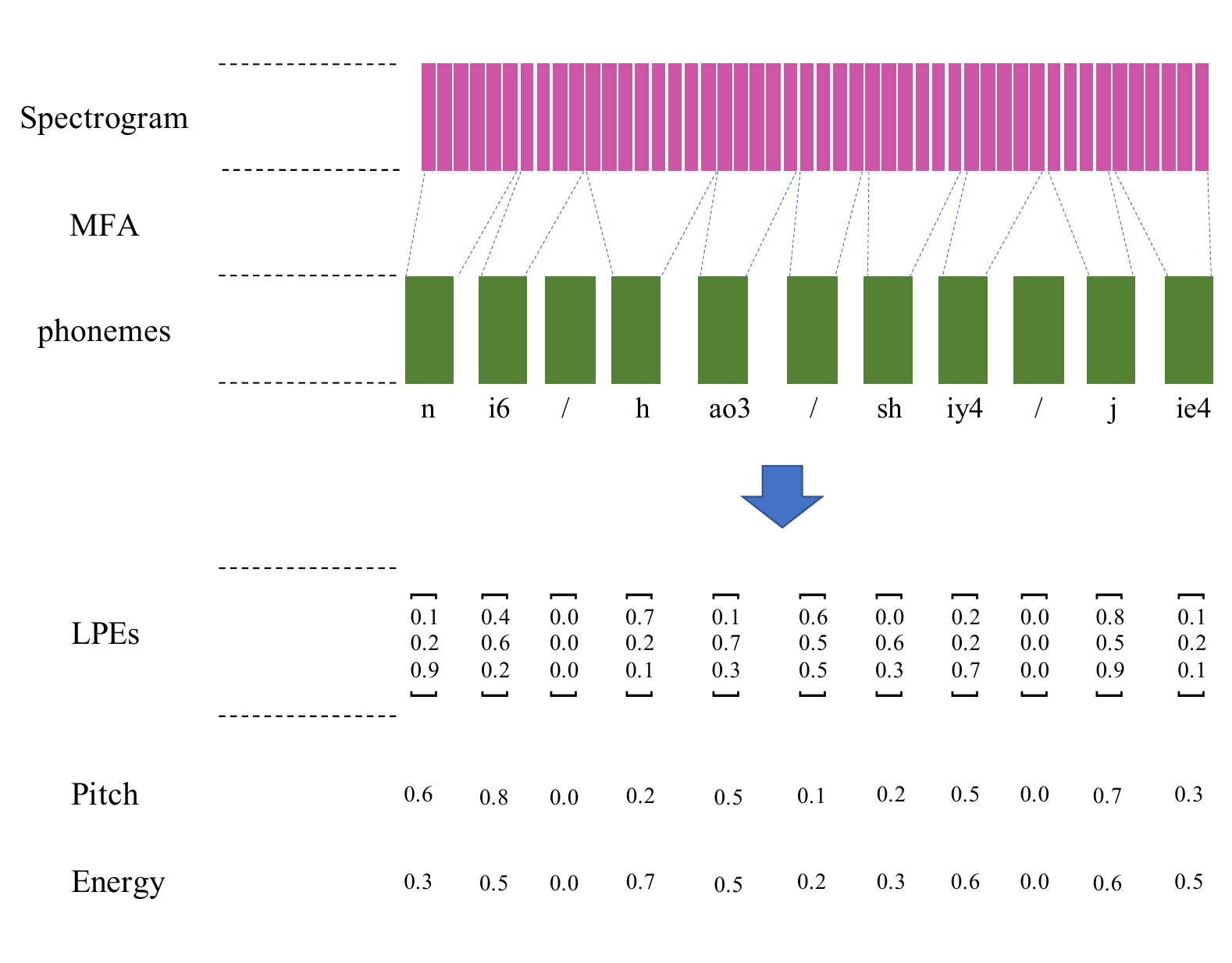}
    \caption{An example of using MFA to align the speech spectrogram and phoneme sequence to extract LPEs features.}
    \label{fig:mfa}
\end{figure}

We hope that the phoneme-level LPE can control the pause and patterns of emphasis of synthesized speech. We add a separator `/' between each word of the text data and regard it as a phoneme to explicitly establish the pause information between the two words. Punctuation marks are not regarded as words but are still used as the input of our text information, so the phoneme length of punctuation is 0, while the phoneme length to the separator `/' is 1. We use Montreal Forced Alignment (MFA)~\cite{MichaelMcAuliffe2017MontrealFA} to extract phoneme boundaries. For each phoneme, we first extract two explicit acoustic features pitch (logF0) and energy, and take the average per frame of each phoneme as the corresponding phoneme-level features. In addition, we extract a phoneme-level 3-dim latent feature (between 0.0 and 1.0) from the speech spectrum, which is combined with the two explicit acoustic features above to reconstruct the original speech. It is worth noting that the pitch and energy information has clear meaning, while the other 3-dim information does not. For the special separator `/' between every two words, if the result extracted by MFA is `silence', the extraction method of the phoneme feature corresponding to `/' is the same as that of the normal pronunciation phoneme. Otherwise, all the phoneme features are 0, which means there is no pause between the two words. Besides, we also jointly extract a GSE from multiple utterances in the same discourse to control the overall style of the synthesized speech. Figure \ref{fig:mfa} shows an example of using MFA to align the speech spectrogram and phoneme sequence to extract LPEs features.

\subsection{Utterance-level Multi-scale Prosodic analysis}
We take the phoneme-level pitch, energy, and 3-dim LPE features obtained in the first subsection as the first objective of our proposed utterance-level model (U-MPM). Below we briefly analyze the information that may be helpful for LPE feature prediction. As we mentioned in related work, exploring prosody features from text is essentially an NLU task, so we first extract word-level embeddings by exploiting the strong representation ability of the pre-trained model for textual information. In addition, prosody is still related to some information at the phoneme level, such as the specific phoneme and tone information. Although we use speaker embedding in the speech synthesis system to control the synthesized speech's timbre, since each speaker's prosody has its characteristics, our LPE information is still inevitably mixed with some speaker-related information. Therefore, we also introduce speaker embedding, style embedding, and dialogue embedding, in which speaker embedding is used to distinguish the prosody features of different speakers, style embedding is used to distinguish the prosody features of different kinds of AudioBooks, and dialogue embedding is used to distinguish whether each word is in dialogue or narration.

Based on the above analyses, we propose U-MPM to predict phoneme-level pitch, energy, and LPE sequences. Our model is shown in the Figure~\ref{fig:3M_sentence}. We use Chinese Roberta~\cite{cui-etal-2020-revisiting} as the word-level text encoder for our model. The resulting word-level features are then added to other word-level features (dialogue embedding) to obtain the final word-level features. Motivated by Fastspeech~\cite{YiRen2019FastSpeechFR}, we use the Length Regulator module to fill the granularity gap between word and phoneme, replicating features at the word level to the phoneme level. The length of the phonemes to each word is known, so we don't need a duration prediction module. The LR module extends the word-level features to the same length as the phoneme. It then fuses with other-level features (phoneme embedding, tone embedding, speaker embedding, style embedding) to obtain the final phoneme-level text representation sequences. Fastspeech2~\cite{YiRen2021FastSpeech2F} uses pitch and energy information to generate more expressive speech, which shows that pitch and energy information can effectively improve prosody expressiveness. We also extract phoneme-level pitch and energy features in the first subsection, and these acoustic features are also used as input during training to improve the accuracy of the predicted LPE. At the same time, our model also predicts these two acoustic features from the text representation sequences.

During training, the model first extracts word embedding from word-level text, and then uses the LR module to expand them to the same length as phonemes. For a utterance $U = (w_{1}, w_{2}, ..., w_{n})$ with separator `/' added, the dialogue embedding is $E_{dia} \in \mathbb{R}^{(n \times d)}$, and its word embedding $E_w \in \mathbb{R}^{(n \times d)}$ is:
\begin{equation}
    E_w = Encoder(w_1, w_2, ..., w_n) + E_{dia}
\end{equation}
where $d$ is the feature dimension of the word embedding. For dialogue labels, we simply treat words in quotation marks as dialogue and other words as narration.
The phoneme length to the word $w_i$ in the utterance is $p_i$. Assuming $p_1=2$, $p_2=1$ (separator `/'), $p_n=0$ (punctuation mark) the length of the expanded word embedding $E_{LR} \in \mathbb{R}^{(N \times d)}$ will be the same as the total length of the phonemes:
\begin{equation}
    E_{LR} = (e_1, e_1, e_2, …, e_{n-1})
\end{equation}
\begin{equation}
    N = \sum_{i=1}^{n}p_{i}
\end{equation}

The phoneme embedding $E_{phn} \in \mathbb{R}^{(N \times d)}$, the tone embedding $E_{tone} \in \mathbb{R}^{(N \times d)}$ and the speaker embedding $E_{spk} \in \mathbb{R}^{d}$ are fused together to get the phoneme feature sequences $E_{pf} \in \mathbb{R}^{(N \times d)}$:
\begin{equation}
    E_{pf} = E_{LR} + E_{phn} + E_{tone} + E_{spk}
\end{equation}
Here we take the tone of each word as the tone of its last phoneme, and the other phonemes' tone labels are all 0.

During training, we use PE predictor to predict acoustic information pitch ($\hat{y}_{pitch}$) and energy ($\hat{y}_{energy}$), and concatenate the real pitch ($y_{pitch} \in\mathbb{R}^{(N \times 1)}$) and energy ($y_{energy} \in\mathbb{R}^{(N \times 1)}$) information with $E_{pf}$ to predict LPE features $\hat{y}_{lpe} \in\mathbb{R}^{(N \times 3)}$  with LPE predictor:
\begin{equation}
    (\hat{y}_{pitch}, \hat{y}_{energy}) = PE\_Predictor(E_{pf})
\end{equation}
\begin{equation}
    E_{t\&a} = concat(E_{pf}, y_{pitch}, y_{energy})
\end{equation}
\begin{equation}
    \hat{y}_{lpe} = LPE\_Predictor(E_{t\&a})
\end{equation}
Both PE Predictor and LPE Predictor are two-layer bidirectional LSTM~\cite{SeppHochreiter1997LongSM} and linear layers combined.

During inference, the acoustic features pitch and energy will be replaced with predicted values instead:
\begin{equation}
    E'_{t\&a} = concat(E_{pf}, \hat{y}_{pitch}, \hat{y}_{energy})
\end{equation}

\begin{equation}
    \hat{y'}_{lpe} = LPE\_Predictor(E'_{t\&a})
\end{equation}

Meanwhile, for an utterance with style label $y$, we use a simple MLP classifier to classify the utterance-level representation (the embedding corresponding to `[CLS]') as the predicted utterance style label $\hat{y}$. During inference, the style embedding can be taken using the predicted style label. Since our speakers are unrelated to the text and cannot be predicted from the text, the speaker labels are specified. Finally, the model can output a 5-dim feature for each phoneme, that is, a 3-dim LPE feature, pitch feature, and energy feature. These features will be provided to the speech synthesis model to synthesize speech. The whole process only needs to specify the speaker and provide the text and pinyin, and the highly expressive speech that conforms to the text characteristics can be generated.

The loss function can be divided into four parts now:
\begin{equation}
    \mathcal{L} = \lambda_{pitch}\mathcal{L}_{pitch} + \lambda_{energy}\mathcal{L}_{energy} + \lambda_{lpe}\mathcal{L}_{lpe} + \lambda_{gse}\mathcal{L}_{gse}
\end{equation}
where the first three parts represent the mean square error (MSE) loss and the last part represents the cross-entropy (CE) loss.

\begin{figure*}
    \centering
    \includegraphics[width=0.9\textwidth]{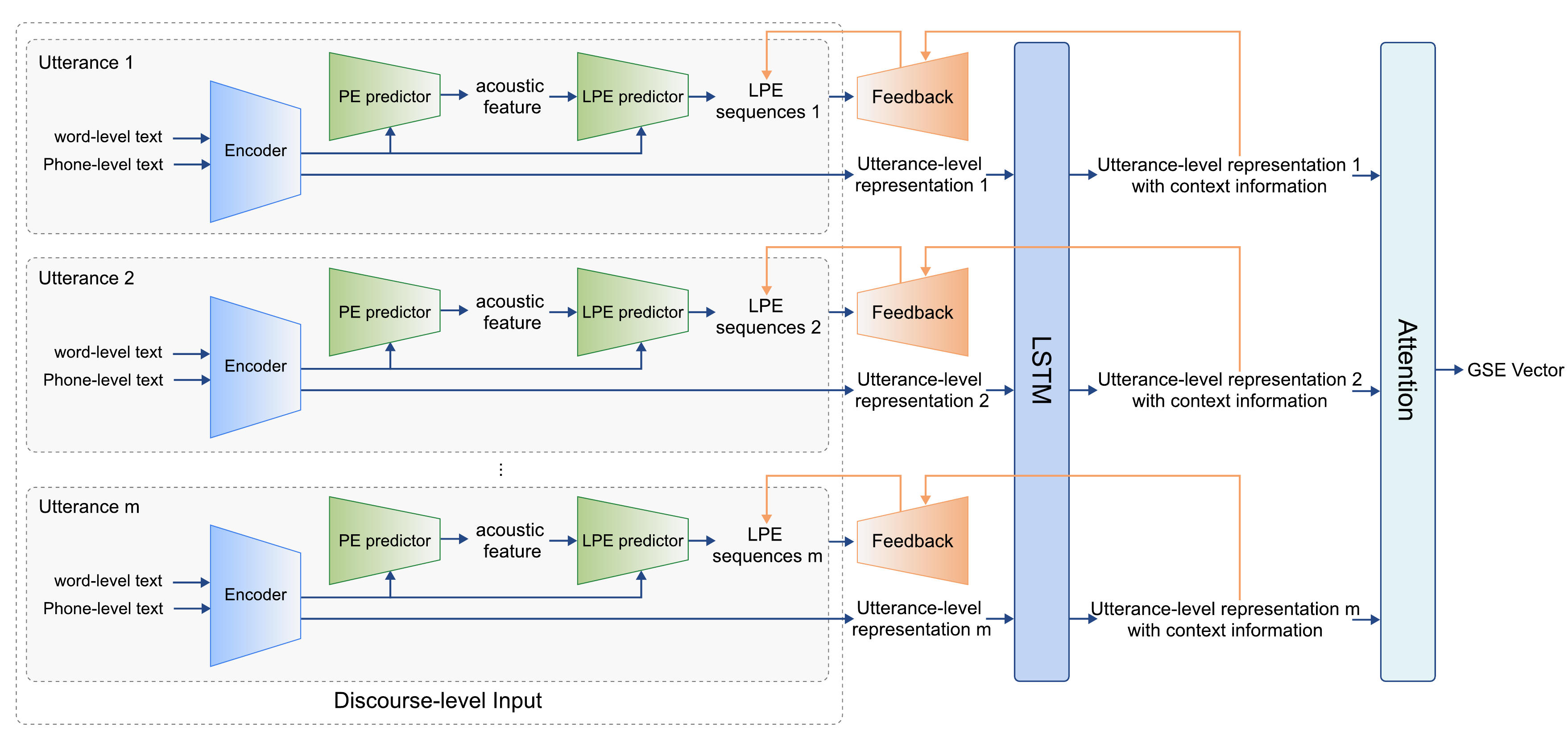}
    \caption{The second stage of our entire training pipeline. At this stage, we freeze the parameters of the single-utterance model trained in the first stage and input all utterance-level features belonging to the same discourse into a bidirectional LSTM to fuse contextual information. Then, the utterance-level features and the LPE feature sequence obtained in the first stage are fed together into $W$, which is a 3D matrix with trainable parameters (convert utterance-level features to phoneme level). $\Delta$LPE is added to the LPE obtained in the first stage to calculate the final LPE features. }
    \label{fig:3M_document}
\end{figure*}

\subsection{Discourse-level Multi-scale Prosodic analysis}
We divide the whole experiment into two stages. The first stage trains the utterance-level model U-MPM in Figure~\ref{fig:3M_sentence}, and the second stage freezes the parameters of the model obtained in the first stage and uses the context information to make certain adjustments to the predicted LPE features. And predict the overall GSE feature of the discourse. Single-utterance-based models cannot exploit discourse-level context. Although the most appropriate prosody of each utterance is mainly determined by the textual content of the utterance itself, different contexts also have certain influences. To make better use of the context of the discourse, based on U-MPM in the previous subsection, we construct a Discourse-level Multi-scale Prosodic analysis Model (D-MPM).

To reflect contextual information in LPE, all phoneme-level features belonging to the same discourse can be fed into an RNN network. However, RNNs perform poorly on very long sequences. Therefore, we treat the output of `[CLS]' of the pre-trained model as utterance-level representation. As shown in Figure~\ref{fig:3M_document}, we feed $m$ utterance-level representations belonging to the same discourse into a bidirectional LSTM to obtain $m$ utterance-level representations fused with contextual information. Then, the previously obtained LPE features are adjusted using these utterance-level features with contextual information. 

Assuming that there are $m$ utterances in a discourse, and the maximum phoneme length in the discourse is $N$, the shape of the predicted LPE feature $\hat{y}_{lpe}$ is (m, N, d), $d=3$, and the shape of the utterance-level feature $U$ is (m, r). We use a matrix $D \in\mathbb{R}^{(r \times d \times d)}$ to map the utterance-level features back to the phoneme level. The advantage of this is that each phoneme can be adjusted more specifically according to utterance-level features and LPE features. Here we use ``einsum'' to complete this calculation, specifically: 
\begin{equation}
    \Delta y_{lpe} = einsum(`rdd,mr,mNd \rightarrow mNd', D, W, \hat{A}_{lpe})
\end{equation}
\begin{equation}
    \hat{y}_{lpe} \mathrel{+}= \Delta y_{lpe}
\end{equation}
Furthermore, the attention mechanism \cite{DzmitryBahdanau2015NeuralMT} is utilized to fuse $m$ utterance-level features into a final discourse-level vector. We still use the same MLP classifier as in stage 1 to predict a global style label.

In this subsection, we get new LPE features as well as a GSE label. The loss function in stage 2 is as follows:
\begin{equation}
    \mathcal{L} = \lambda_{lpe}\mathcal{L}_{lpe} + \lambda_{gse}\mathcal{L}_{gse}
\end{equation}
The loss function consists of three parts: the first part is the MSE loss of the adjusted LPE features, the second part is the CE loss of the discourse style labels.

\section{Experiments}
In this section, we conduct extensive experiments to demonstrate the effectiveness of our framework. We evaluate our framework on the collected Discourse-level Chinese AudioBook dataset (DCA). We compared with models that only use word-level features, only use phoneme-level features, and do not combine acoustic features (pitch and energy) with text features to demonstrate the effectiveness of our proposed multi-grained and multi-modal input. We also compare the difference between U-MPM and D-MPM to prove the improvement of using discourse data on prosody prediction. Our model can effectively predict local and global prosody information from discourse-level text. Ultimately, these prosody features can be used by a speech synthesis system to synthesize speech that is consistent with the style of the text.

\subsection{Dataset}
As we mentioned in the introduction section, we build a Discourse-level Chinese AudioBook dataset (DCA). We collected two styles of books, one is fairy tale style and the other is martial arts novel style. The first style is mainly aimed at children, while the second is mainly aimed at groups of teenagers or adults. We divide these stories into discourse-level paragraphs, and each discourse-level text contains an average of about 10 utterances. So far, our DCA includes 10 speakers of different genders and age groups. In total, it contains 1,913 discourse-level data, or 19,988 utterances. Among them, the data with pinyin label (Chinese pronunciation notation) has been marked with 8 speakers, with a total of 13,116 utterances. We use 12,341 utterances as our training set and the remaining 775 utterances as our test set. Both the training set and the test set are divided into discourse. 
% The statistical results of the labeled data in DCA can be seen in Table~\ref{tab:dataset}.

% \begin{table}[htbp]
% \caption{Statistics labeled data in our DCA. The `f' in gender represents female and `m' represents male. The `C' in age stands for childhood, `Y' for youth, `M' for middle age, and `O' for old age. `FT' in the style stands for fairy tale and `MA' stands for martial arts novel.}
% \begin{tabular}{c|ccccc}
% % \toprule[1.5pt]
% \hline
% \textbf{Speaker} & \textbf{Gender} & \textbf{Age} & \textbf{Style} & \multicolumn{1}{l}{\textbf{Discourse}} & \textbf{Utterance} \\
% \hline
% \hline
% \rule{0pt}{10pt}
% \textbf{01} & f & Y & FT & 177 & 2069 \\
% \hline
% \rule{0pt}{10pt}
% \textbf{02} & m & Y & FT & 213 & 2795 \\
% \hline
% \rule{0pt}{10pt}
% \textbf{02} & m & Y & MA & 329 & 2808 \\
% \hline
% \rule{0pt}{10pt}
% \textbf{03} & f & C & FT & 91 & 872 \\
% \hline
% \rule{0pt}{10pt}
% \textbf{04} & f & M & FT & 42 & 525 \\
% \hline
% \rule{0pt}{10pt}
% \textbf{04} & f & M & MA & 57 & 504 \\
% \hline
% \rule{0pt}{10pt}
% \textbf{05} & m & C & FT & 88 & 1128 \\
% \hline
% \rule{0pt}{10pt}
% \textbf{07} & f & Y & FT & 117 & 1213 \\
% \hline
% \rule{0pt}{10pt}
% \textbf{10} & m & M & MA & 59 & 662 \\
% \hline
% \rule{0pt}{10pt}
% \textbf{12} & f & O & FT & 47 & 540 \\
% \hline
% % \bottomrule[1.5pt]
% \end{tabular}
% \label{tab:dataset}
% \end{table}

\subsection{Implementation Details}
Our model training consists of two stages in total: the first stage trains the utterance-level model mentioned in Section 4.2. In the second stage, we freeze the model parameters trained in the first stage, and on this basis, continue to use the discourse-level data to adjust the LPE features and obtain the GSE label.

We train our model on one Geforce RTX 1080Ti GPU. We set different learning rates for different parts of the model. In the first stage, we set a learning rate of 1e-5 to finetune the pre-trained Chinese Roberta~\cite{cui-etal-2020-revisiting}, and a learning rate of 1e-3 for the rest. In the second stage, we freeze the parameters trained in the first stage and set a learning rate of 2e-4 for the newly added modules. The input of the first stage is in utterance units, and we train with a mini-batch size of 16 using Adam optimizer~\cite{DiederikPKingma2014AdamAM}. The input of the second stage is in discourse units, and the min-batch size is set to 32. We train our model for 40 epochs at each stage (all the models have converged). The pitch (logF0) and energy features we use during training are quite different from the LPE feature scale, so we give different weights to the coefficients of different parts of the loss function. Specifically, $\lambda_{pitch}$, $\lambda_{energy}$, $\lambda_{lpe}$, $\lambda_{gse}$ are respectively 0.05, 0.0025, 1.0 and 1.0. We use Tacotron2~\cite{JonathanShen2018NaturalTS} as the backbone of the speech synthesis module and LPCNET~\cite{JeanMarcValin2019LPCNETIN} as our vocoder.

\begin{table*}
\caption{MOS Evaluation Result. PE: pitch and energy.}
\label{tab:mos}
\centering
\setlength{\tabcolsep}{2.5mm}
\begin{tabular}{c|cccc}
% \toprule[1.5pt]
\hline
\textbf{Method} & \textbf{Naturalness of Pauses} & \textbf{Expressiveness of Rhythm and Emphasis} & \textbf{Naturalness and Coherence} & \textbf{LPEs MSE}\\
\hline
\hline
\rule{0pt}{10pt}
U-MPM & 4.175 $\pm$ 0.701  & 4.078 $\pm$ 0.580 & 3.948 $\pm$ 0.737 & 0.02383 \\
\hline
\rule{0pt}{10pt}
w/o word & 1.987 $\pm$ 0.734 & 2.714 $\pm$ 0.681 & 2.279 $\pm$ 0.793 & 0.03652 \\
\hline
\rule{0pt}{10pt}
w/o phn & 4.026 $\pm$ 0.765 & 3.766 $\pm$ 0.742 & 3.805 $\pm$ 0.722 & 0.02597 \\
\hline
\rule{0pt}{10pt}
w/o PE & \textbf{4.240 $\pm$ 0.672} & 3.942 $\pm$ 0.628 & 3.987 $\pm$ 0.683 & \textbf{0.02172} \\
\hline
\rule{0pt}{10pt}
transfer & 3.742 $\pm$ 0.876 & \textbf{4.152 $\pm$ 0.619} & 3.910 $\pm$ 0.805 & - \\
\hline
D-MPM & 4.182 $\pm$ 0.702 & 4.123 $\pm$ 0.624 & \textbf{4.110 $\pm$ 0.588} & 0.02257 \\
\hline
% \bottomrule[1.5pt]
\end{tabular}
\end{table*}

\subsection{Compare Different Model Structures}
Our proposed text analysis model focuses on LPE as targets. However, LPE is a latent feature extracted by the reference encoder, and it is difficult to intuitively judge its performance. In the experiments, we take the MSE loss of the LPE sequence as the main component of the objective function. The style and prosody of speech are relatively subjective concepts. Therefore, we use the mean opinion score (MOS) to evaluate the synthesis results. We select 7 fragments from the test set and outside and invit 11 participants to rate the synthesized speech from three perspectives: (1) the Naturalness of Pauses; (2) the Expressiveness of Rhythm and Emphasis; (3) the style Naturalness and Coherence of the whole discourse. Participants can rate in the range of 1 to 5, with higher scores indicating greater expressiveness. 

The results are shown in Table~\ref{tab:mos}. We remove part of the input based on U-MPM and retrained the model holding other conditions constant to explore the influence of different features on the final prosody. We remove word-level features (w/o word), phoneme-level features (w/o phn), and acoustic features (w/o PE) and \textbf{retrain} these models to compare with U-MPM.

It can be observed that using only phoneme-level text as input (w/o word), the model trained in our dataset with a not huge amount of data but rich prosody has a poor effect, especially in the prediction of pause. In contrast, U-MPM with multi-scale features significantly improved in all three indicators (+2.188, +1.364, +1.669), suggesting that word-level analysis could help the model improve reading comprehension, thus better-predicting prosody style features and generating more expressive speech. Similarly, if we remove all the phone-level information (w/o phn), although the score does not decrease as significantly as that of word-level information, it still decreases somewhat, especially in prosody expressiveness (-0.312). We observed that when phoneme-level information is removed, the intonation of some sounds became less natural and the overall sense of rhythm is weakened. A very interesting phenomenon is that if the acoustic features (pitch and energy) prediction step is removed (w/o PE), that is, directly predicting the LPE sequence from the fusion of multi-grained text information, we find that the LPE loss function is the minimum, and it also performs best in the pause naturalness. However, the prosody expressiveness of synthetic speech is significantly reduced (-0.136). We observed that there is a weak sense of change in stress in the synthesized speech. We believe that the addition of acoustic modes makes the model more aggressive in predicting prosody. Although it may be inconsistent with the target prosody, the model has a strong sense of rhythm, even at the cost of a small amount of pause naturalness. However, the prediction of LPE without acoustic modes is more conservative, which makes the loss function smaller, but the sense of rhythm is lost. 

Another interesting phenomenon is that our text analysis model uses the features extracted from the style transfer model as target (which is equal to LPE MSE Loss is 0), but the final result is significantly better than the speech synthesized by the style transfer model in the pause naturalness (+0.433). This is because when a person reads a text paragraph, there may be a certain amount of randomness, which causes the pause to be not very strict. However, the results obtained by the text analysis model are learned from a large amount of data, making the results more stable. We take the prosody features obtained by style transfer model as the target, and finally use the prosody features predicted by text analysis model to guide the real synthetic scene, which can get more expressive speech than style transfer. This is why our LPE MSE Loss is not the lower the better, but can only be used as a soft constraint. The overall performance of our proposed D-MPM is the best. Compared with the U-MPM, D-MPM can use the context information to adjust the LPE features to guide the global style, which makes the D-MPM model +0.007, +0.045, +0.162 on the three indicators, respectively. It can be seen that adding the training task of the second stage and training the entire discourse together can make each utterance not only pay attention to itself but also its context, making the overall style more natural and coherent. The third indicator measures the overall naturalness and coherence of synthesized speech, which is extremely important for discourse-level synthesis tasks such as AudioBooks. Compared with U-MPM that only uses single utterance training, D-MPM adds the whole discourse training stage, which makes great progress in its comprehensive performance.

\begin{figure}
    \centering
    \subfigure[U-MPM]{\includegraphics[width=0.47\columnwidth]{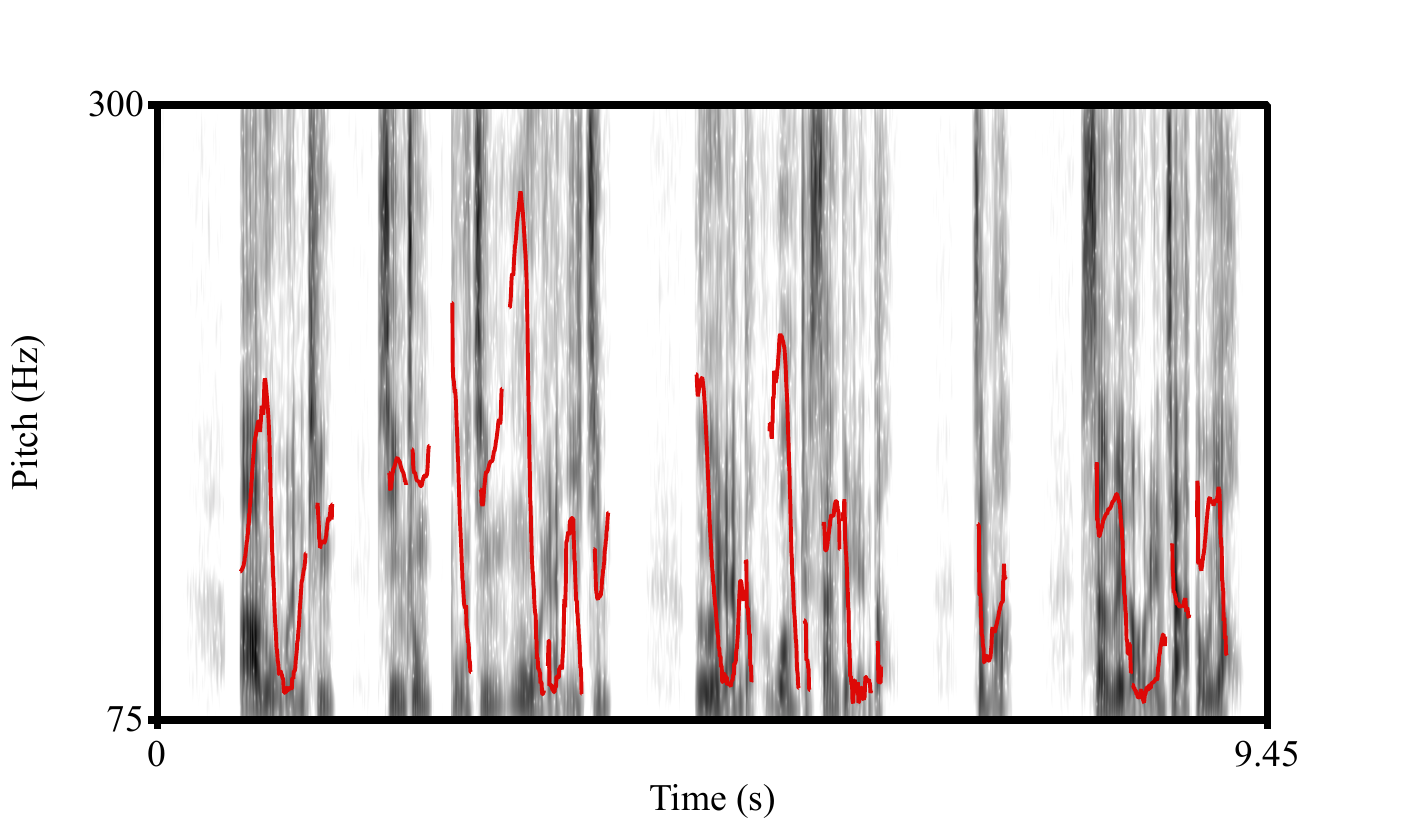}}
    \quad
    \subfigure[w/o word]{\includegraphics[width=0.45\columnwidth]{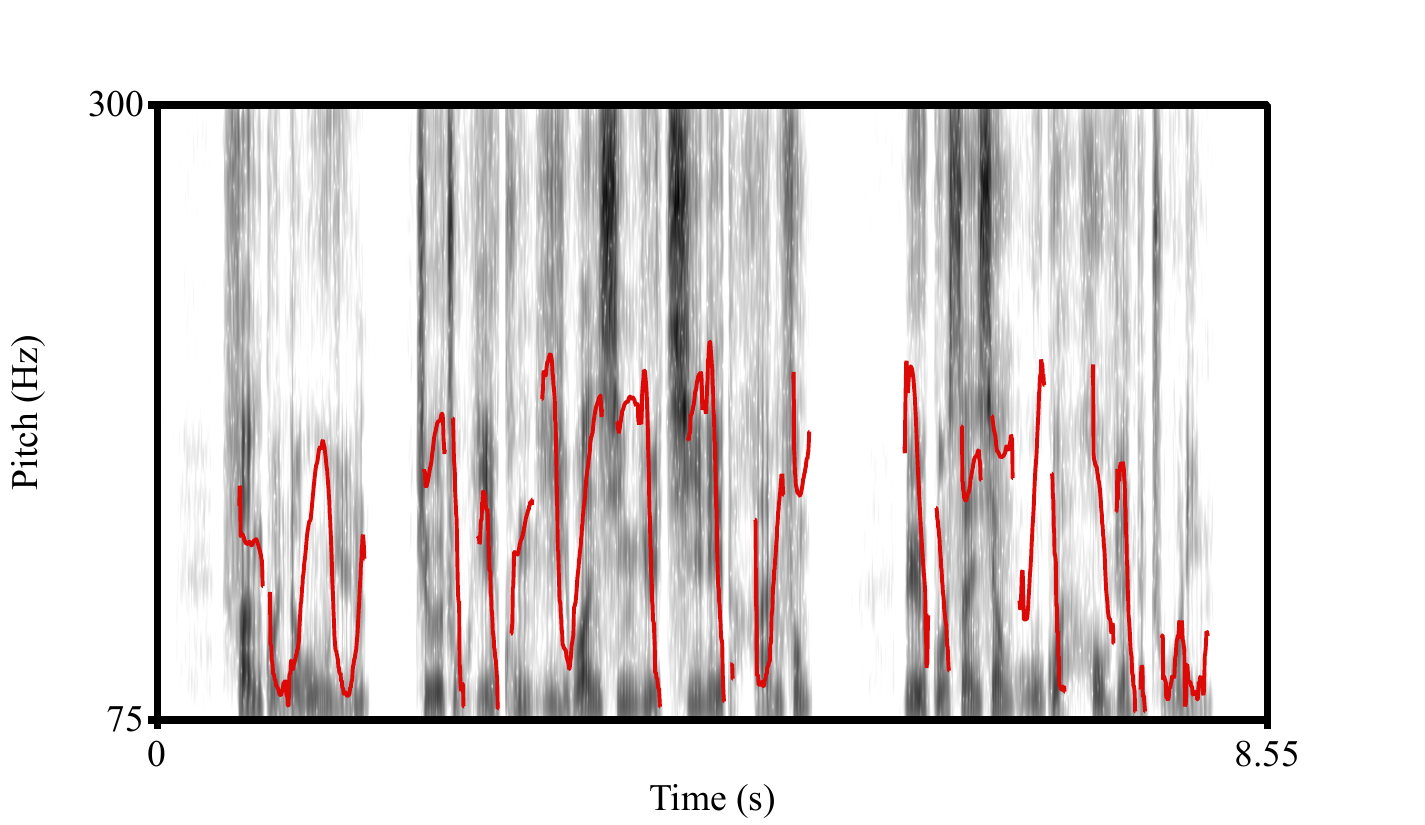}}
    
    \subfigure[w/o acoustic]{\includegraphics[width=0.45\columnwidth]{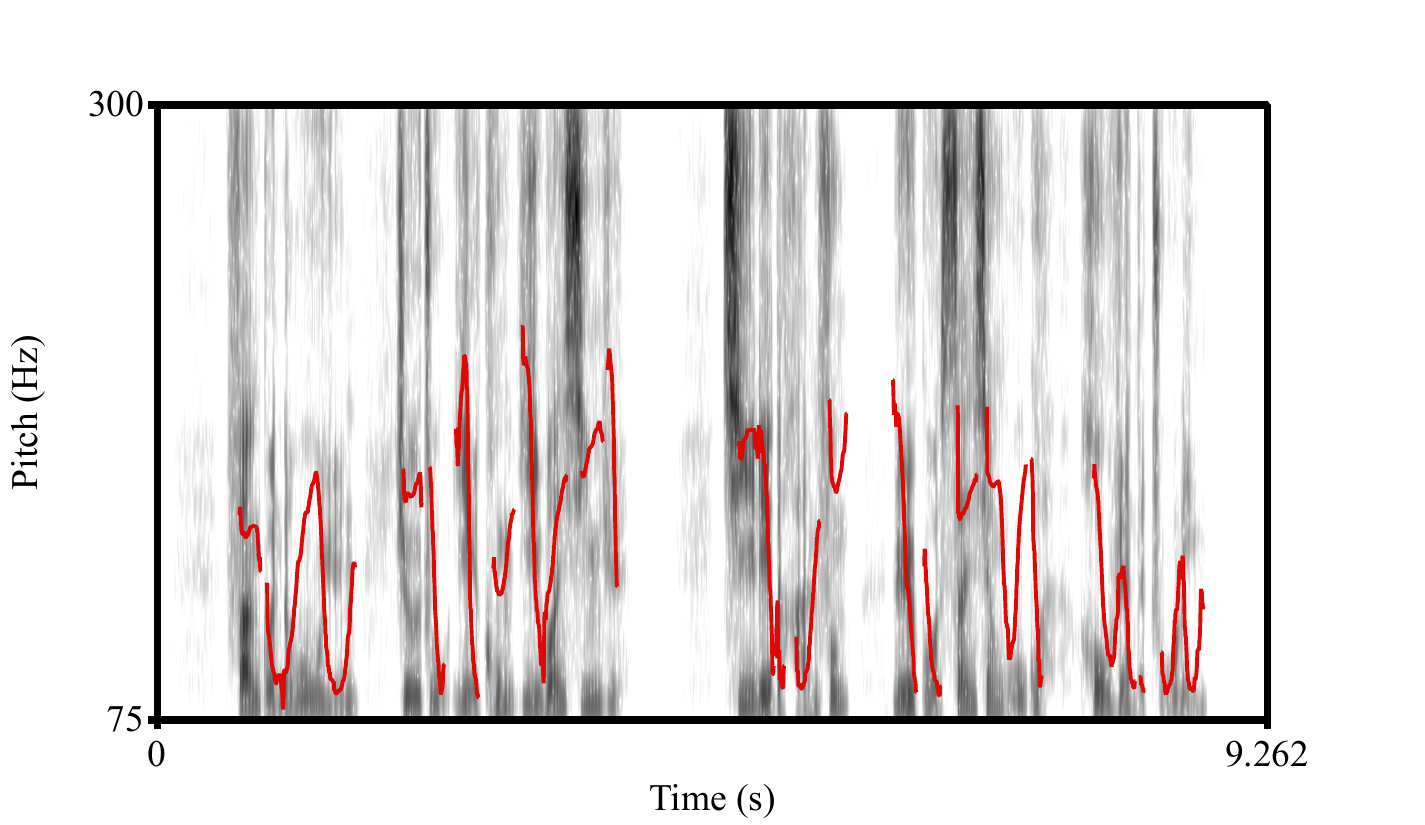}}
    \quad
    \subfigure[D-MPM]{\includegraphics[width=0.45\columnwidth]{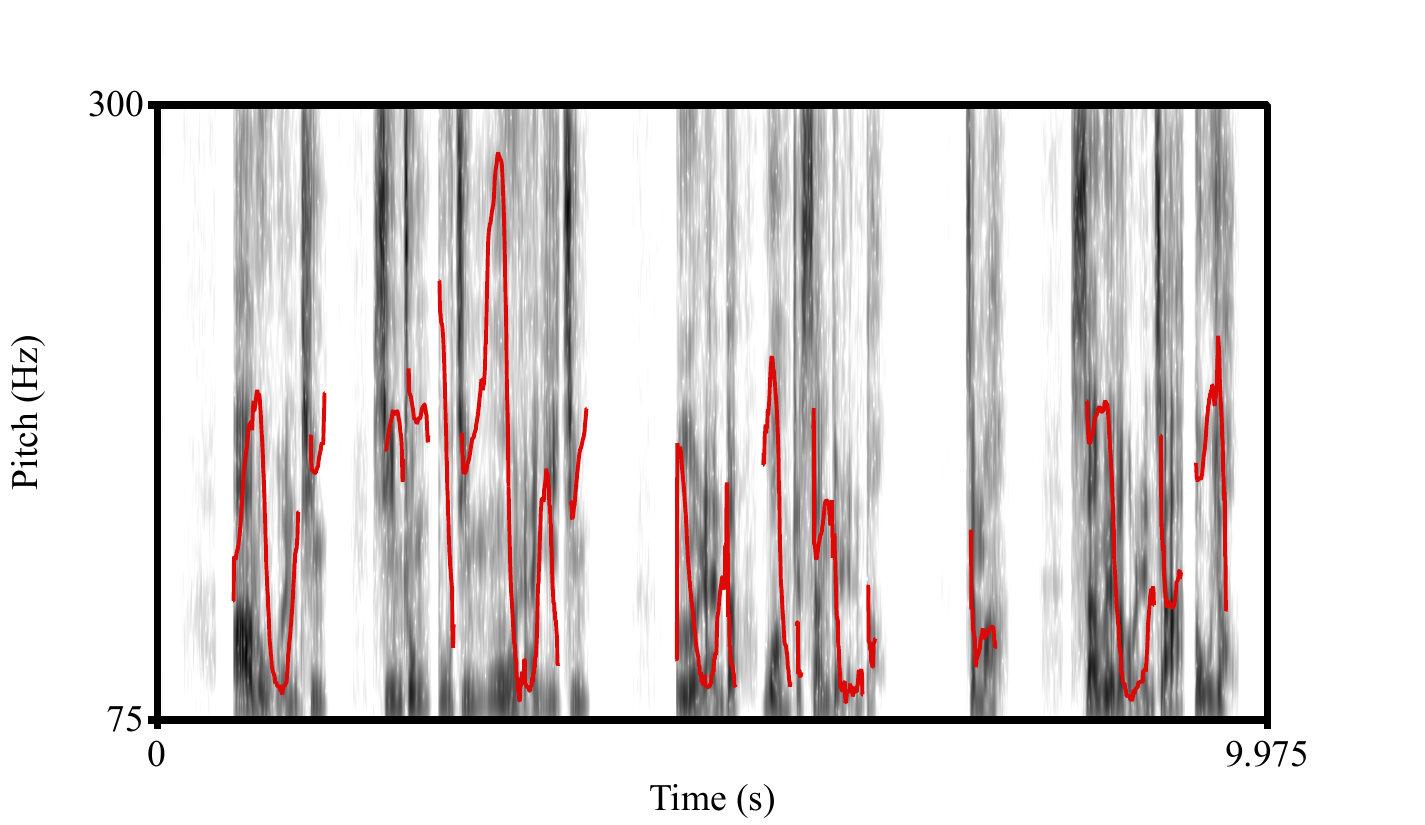}}
    \caption{The variation trend of pitch contours of the same speech synthesized by different models.}
    \label{fig:pitch}
\end{figure}

\subsection{Case Study}
To visualize the prosody features of synthesized speech, we use the prosody features obtained by four different models to control the speech synthesis system to synthesize speech and use Praat~\cite{ZhongqiangHuang2006AnOS} to draw the pitch contour of the same content synthesized by different models. As can be seen from Figure~\ref{fig:pitch}, the pause of the U-MPM w/o word model is significantly different from the other three models, and there is a longer part in the middle without pause, which sounds very unnatural. This also corresponds to the poor performance of U-MPM w/o word on pause naturalness in Table~\ref{tab:mos}. While the U-MPM w/o acoustic model only uses textual information to predict prosody features without resorting to intermediate acoustic features. It can be seen that the speech synthesized by the model without the fusion of acoustic features appears to be relatively monotonous in the change of pitch, and the range of pitch change is small. There is no obvious difference between U-MPM and D-MPM, indicating that the context only plays a small adjustment role, and the content of the utterance itself still plays a decisive role. However, the global style and coherence of D-MPM are better. In a discourse-level speech synthesis task like AudioBook synthesis, our proposed D-MPM can achieve overall better results.

\section{Conclusion}
In this paper, we release a new Discourse-level Chinese Audiobook (DCA) dataset and propose a novel discourse-level multi-scale text prosodic analysis model. The model takes the phoneme-level fine-grained latent prosodic feature embedding, local prosody embedding (LPE), extracted by a style transfer model as the main target. Experimental results show that the proposed method can effectively utilize the information from different scales of text to synthesize more expressive speech. And the introduction of discourse context information can make the final synthesized speech have a better global coherence performance. 

\bibliographystyle{IEEEtran}
\bibliography{chinamm}

\end{document}